# AN ALGORITHM FOR ODD GRACEFULNESS OF THE TENSOR PRODUCT OF TWO LINE GRAPHS

M. Ibrahim Moussa

Faculty of Computers & Information, Benha University, Benha, Egypt
moussa_6060@yahoo.com

## ABSTRACT

*An odd graceful labeling of a graph $G = (V, E)$ is a function $f : V(G) \to \{0,1,2, \ldots 2|E(G)|-1\}$ such that $|f(u)-f(v)|$ is odd value less than or equal to $2|E(G)|-1$ for any $u,v \in V(G)$. In spite of the large number of papers published on the subject of graph labeling, there are few algorithms to be used by researchers to gracefully label graphs. This work provides generalized odd graceful solutions to all the vertices and edges for the tensor product of the two paths $P_n$ and $P_m$ denoted $P_n \wedge P_m$. Firstly, we describe an algorithm to label the vertices and the edges of the vertex set $V(P_n \wedge P_m)$ and the edge set $E(P_n \wedge P_m)$ respectively. Finally, we prove that the graph $P_n \wedge P_m$ is odd graceful for all integers $n$ and $m$.*

## KEY WORDS
*Vertex labeling, edge labeling, odd graceful, Algorithms.*

## 1. INTRODUCTION

Let $G$ is a finite simple graph, whose vertex set denoted $V(G)$, and the edge set denoted $E(G)$ The order of $G$ is the cardinality $n = |V(G)|$ and the size of $G$ is the cardinality $q = |E(G)|$. We write $uv \in E(G)$ if there is an edge connecting the vertices $u$ and $v$ in $G$. A path graph $P_n$ simply denotes the graph that consists of a single line. In other words, it is a sequence of vertices such that from each of its vertices there is an edge to the next vertex in the sequence. The first vertex is called the start vertex and the last vertex is called the end vertex. Both of them are called end or terminal vertices of the path. The other vertices in the path are internal vertices. The tensor product of two graphs $G_1$ and $G_2$ denoted $G_1 \wedge G_2$ its vertex set denoted $V(G_1 \wedge G_2) = V(G_1) \times V(G_2)$ consider any two points $u = (u_1, u_2)$, and $v = (v_1, v_2)$ in $V(G_1 \wedge G_2)$ the edge set denoted $E(G_1 \wedge G_2)$ where $E(G_1 \wedge G_2) = \{(u_1,v_1)(u_2,v_2) : u_1u_2 \in E(G_1) \ \& \ v_1v_2 \in E(G_2)\}$

The tensor product is also called the direct product, or conjunction. The tensor product was introduced by Alfred North Whitehead and Bertrand Russell in their Principia Mathematica [3]. The graph $G = (V, E)$ consists of a set of vertices and a set of edges. If a nonnegative integer $f(u)$ is assigned to each vertex $u$, then the vertices are said to be "labeled." $G = (V, E)$ is itself a labeled graph if each edge $e$ is given the value $f^*(e) = |f(u) - f(v)|$ where $u$ and $v$ are the endpoints of $e$. Clearly, in the absence of additional constraints, every graph can be labeled in infinitely many ways. Thus utilization of labeled graph models requires imposition of additional constraints which characterize the problem being investigated.

An odd graceful labeling of the graph $G$ with $n = |V(G)|$ vertices and $q = |E(G)|$ edges is a one-to-one function $f$ of the vertex set $V(G)$ into the set $\{0, 1, 2, ..., 2q-1\}$ with this property: if we define, for any edge $uv$ the function $f^*(uv) = |f(u) - f(v)|$ the resulting edge label are $\{1, 3, ..., 2q-1\}$. A graph is called odd graceful if it has an odd graceful labeling. The odd graceful labeling problem is to find out whether a given graph is odd graceful, and if it is odd graceful, how to label the vertices. The common approach in proving the odd gracefulness of special classes of graphs is to either provide formulas for odd gracefully labeling the given graph, or construct desired labeling from combining the famous classes of odd graceful graphs.

The study of odd graceful graphs and odd graceful labeling definition was introduced by Gnanajothi [9], she proved the following graphs are odd graceful: the graph $C_m \times K_2$ is odd-graceful if and only if $m$ even, and the graph obtained from $P_n \times P_2$ by deleting an edge that joins to end points of the $P_n$ paths, this last graph knew as the ladder graph. She proved that every graph with an odd cycle is not odd graceful. This labeling has been studied in several articles. In 2010 Moussa [19] have presented the algorithms that showed the graph $P_n \cup C_m$ is odd graceful if $m$ is even. For further information about the graph labeling, we advise the reader to refer to the brilliant dynamic survey on the subject made by J. Gallian in his dynamic survey [18].

In this paper, we first explicitly defined an odd graceful labeling of $P_n \wedge P_2$, $P_n \wedge P_3$, $P_n \wedge P_4$ and $P_n \wedge P_5$ then, using this odd graceful labeling, described a recursive procedure to obtain an odd graceful labeling of the graph $P_n \wedge P_m$. Finally; we presented an algorithm for computing the odd graceful labeling of the tensor graph $P_n \wedge P_m$, the correctness of the algorithm was proved at the end of this paper. The remainder of this paper is organized as follows. In section 2, we gave a nice set of applications related with the odd graceful labeling. In section 3 defined the tensor graph of two path graphs, and then drew the tensor graph $P_n \wedge P_m$ on the plane. Section 4 described the algorithm of odd graceful labeling of the tensor graph $P_n \wedge P_m$ and then proved the correctness of the algorithm. The computations cost is computed at the end of this section. Section 5 is the conclusion of this research.

## 2. THE APPLICATION RANGES

The odd graceful labeling is one of the most widely used labeling methods of graphs. While the labeling of graphs is perceived to be a primarily theoretical subject in the field of graph theory and discrete mathematics, it serves as models in a wide range of applications.

The coding theory: The design of certain important classes of good non periodic codes for pulse radar and missile guidance is equivalent to labeling the complete graph in such a way that all the edge labels are distinct. The node labels then determine the time positions at which pulses are transmitted.

The x-ray crystallography: X-ray diffraction is one of the most powerful techniques for characterizing the structural properties of crystalline solids, in which a beam of X-rays strikes a crystal and diffracts into many specific directions. In some cases more than one structure has the same diffraction information. This problem is mathematically equivalent to determining all labeling of the appropriate graphs which produce a prespecified set of edge labels.

The communications network addressing: A communication network is composed of nodes, each of which has computing power and can transmit and receive messages over communication links, wireless or cabled. The basic network topologies are include fully connected, mesh, star, ring, tree, bus. A single network may consist of several interconnected subnets of different topologies. Networks are further classified as Local Area Networks (LAN), e.g. inside one building, or Wide Area Networks (WAN), e.g. between buildings. It might be useful to assign each user terminal a "node label," subject to the constraint that all connecting "edges" (communication links) receive distinct labels. In this way, the numbers of any two communicating terminals automatically specify (by simple subtraction) the link label of the connecting path; and conversely, the path label uniquely specifies the pair of user terminals which it interconnects [10].

Cellular networks are most successful commercial application of wireless networks. In a cellular network, a service coverage area is divided into smaller square or hexagonal areas referred to as cells. Each cell is served by a base station. The base station is able to communicate with mobile stations such as cellular telephones, using its radio transceiver. A mobile switching center (MSC) connects the base station with the public switched telephone network (PSTN). There are two possible cell configurations to cover the service area, the first is a hexagonal configuration which is commonly used where each cell has six neighboring cells and the second is a mesh configuration which is more simple where four neighbors can be assumed for each cell (horizontal and vertical ones) [11, 12, 13, 14]. Mesh nets are considered good models for large-scale networks of wireless sensors that are distributed over a geographic region. Many cellular systems operate under constraints. One of these constraints is intermodulation constraint, which interferes with the use of frequency gaps that are multiples of each other [15]. The channel assignment under intermodulation constraint is related to graceful labeling of graphs [10]. The challenge concerning channel assignment is to give maximal channel reuse without violating the constraints so that blocking is minimal [16]. Figure 0 illustrates a tensor graph $P_n \wedge P_m$ with each vertex in the set $V(P_n \wedge P_m)$ represents a base station in a cell and the edges represent geographical adjacency of cells. The black dots in this figure are the vertices of the tensor graph, which correspond to the base stations, and the connected black lines are the edges of the tensor graph, which correspond to adjacent cells in the network.

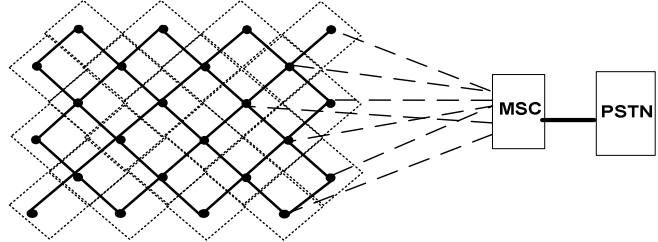

Figure 0. Two-dimensional network topology with the mesh configuration

Radio Channel Assignments: The vertex-labeling of graphs with nonnegative integers provides a natural setting to study problems of radio channel assignment. The frequency spectrum allocated to wireless communications is very limited, so the cellular concept was introduced to reuse the frequency. Each cell is assigned a certain number of channels. To avoid radio interference, the channels assigned to one cell must be different from the channels assigned to its neighboring cells. However, the same channels can be reused by two cells that are far apart such that the radio interference between them is acceptable. By reducing the size of cells, the cellular network is able to increase its capacity, and therefore to serve more subscribers. The notation offset is needed when it uses the graph labeling to find channel assignments to forbid repeated labeling. In 2000 Chartrand, Erwin, Harary, and Zang described the graph labelings with specific constraints related to the diameter of the graph are called radio labelings [17]. On the other side, some of the algorithms for channel assignment with general constraints are based on graph labeling. For further information about the algorithms for channel assignment with general constraints, we advise the reader to go to [10]

For further information about other applications of labeled graphs, such applications include radar, circuit design, astronomy, data base management, on automatic drilling machine, determining configurations of simple resistor networks and models for constraint programming over finite domains (see e.g., [1, 4, 5, 7 and 8]). Labeled graph also apply to other areas of mathematics (see e.g., [2, 6]).

## 3. THE TENSOR OF TWO PATHS

The tensor graph $P_n \wedge P_m$ has a total number of vertices equals $nm$ and $2(n-1)(m-1)$ edge. Let $l_1, l_2, \ldots, l_m$ be $m$ distinct parallel lines lying in a plane. Let $s_1, s_2, \ldots, s_m$ be infinite set of points on $l_1, l_2, \ldots, l_m$ respectively. We consider the vertex set of the graph $P_n \wedge P_m$ is partitioned into vertex sets $V_1, V_2, \ldots, V_m$ drawn so that the vertices lie on distinct consecutive points of $s_1, s_2, \ldots, s_m$ respectively. An arbitrary number $j = 1, 2, \ldots, m$ the set $s_j$ has the vertices $v_i^j$ for all $i = 1, 2, \ldots, n$, this representation shown in Figure 1.

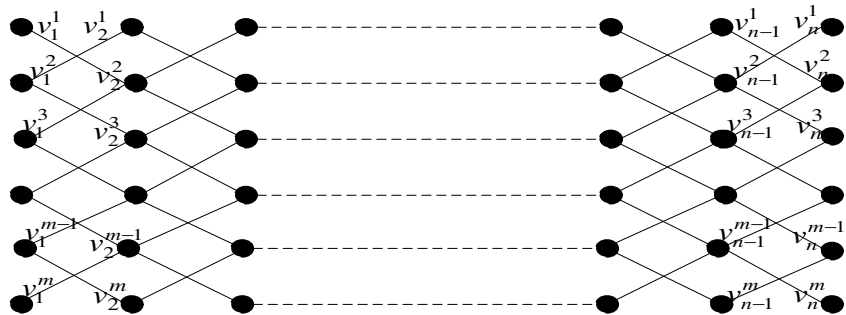

Figure 1 the graph $P_n \wedge P_m$

## 4. VARIATION OF ODD GRACEFUL LABELING OF $P_n \wedge P_m$

### 4.1. Odd Gracefulness of $P_n \wedge P_2$

**Theorem 1** $P_n \wedge P_2$ is odd graceful for every integer $n > 2$.

**Proof**

As shown in the above section, let the vertex set of the graph $P_n \wedge P_2$ is partitioned into two disjoint sets $V_1$ and $V_2$ and drawn so that the vertices lie on distinct consecutive points of $s_1$ and $s_2$ respectively. The total number of vertices equals $2n$ and total number of edges equals $2(n-1)$. For any vertex $v_i^1 \in s_1$ and $v_i^2 \in s_2$ the odd graceful labeling functions $f(v_i^1): s_1 \to \{0, 1, 2, \ldots, 4n-3\}$ and $f(v_i^2): s_2 \to \{0, 1, 2, \ldots, 4n-3\}$ defined respectively by

$$f(v_i^1) = \begin{cases} i - 1 & i \text{ odd} \\ 2n - i & i \text{ even} \end{cases} \qquad f(v_i^2) = \begin{cases} i & i \text{ odd} \\ (4n - 3) - i & i \text{ even} \end{cases}$$

The weights of the vertices in the set $s_1$ belong to the even numbers 0, 2, 4,…,4(*n*-1)-2 and the weights of the vertices in the set $s_2$ belong to the odd numbers 1, 3, 5,…,4(*n*-1)-1. The edge's labeling induced by the absolute value of the difference of the vertex's labeling. The reader can easily find out, that the vertex labels are distinct and all the edge labels are distinct odd labels. Thus, $f$ is odd graceful labeling of $P_n \wedge P_2$. Figure2 shows the method labeling of the graph $P_8 \wedge P_2$ this complete the proof.∎

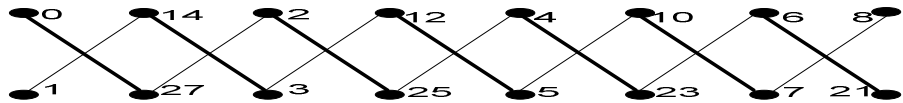

Figure 2 the graph $P_8 \wedge P_2$

### 4.2. Odd gracefulness of $P_n \wedge P_3$

**Theorem 2** $P_n \wedge P_3$ is odd graceful for every integer $n > 2$.

**Proof**

The vertex set of the graph $P_n \wedge P_3$ is partitioned into three disjoint sets $V_1, V_2$ and $V_3$ drawn so that the vertices lie on distinct consecutive points of $s_1, s_2$ and $s_3$ respectively. The total number of vertices equals $3n$ and total number of edges equals $4(n-1)$. For any vertex $v_i^1 \in s_1, v_i^2 \in s_2$ and $v_i^3 \in s_3$ the odd graceful labeling functions $f(v_i^1), f(v_i^2)$ and $f(v_i^3)$ defined respectively by

$$f(v_i^1) = \begin{cases} i-1 & i \text{ odd} \\ (6n-4)-i & i \text{ even} \end{cases} \quad f(v_i^2) = \begin{cases} i & i \text{ odd} \\ (8n-7)-i & i \text{ even} \end{cases} \quad f(v_i^3) = \begin{cases} (4n-5)+i & i \text{ odd} \\ 2n-i & i \text{ even} \end{cases}$$

The weights of the vertices in the set $s_1$ and $s_3$ are the even numbers 0, 2, 4,…, 8(*n*-1)-2 and the weights of the vertices in the set $s_2$ are the odd numbers 1, 3, 5,…,8(*n*-1)-1. The edge's labeling induced by the absolute value of the difference of the vertex's labeling. The reader can easily find out, that the vertex labels are distinct and all the edge labels are distinct odd labels. Thus, $f$ is odd graceful labeling of $P_n \wedge P_3$. Figure3 shows the method labeling of the graph $P_8 \wedge P_3$ this complete the proof.∎

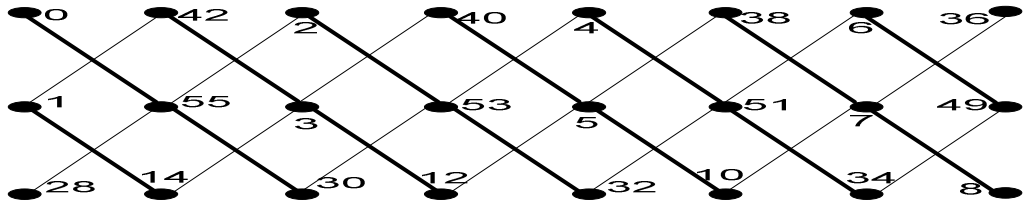

Figure 3 the graph $p_8 \wedge p_3$

## 4.3. Odd gracefulness of $P_n \wedge P_4$

**Theorem 3** $P_n \wedge P_4$ is odd graceful for every integer $n > 2$.

**Proof**:

The vertex set of the graph $P_n \wedge P_4$ is partitioned into four disjoint sets $V_1, V_2, V_3$ and $V_4$ drawn so that the vertices lie on distinct consecutive points of $s_1, s_2, s_3$ and $s_4$ respectively. The total number of vertices equals $4n$ and total number of edges equals $6(n-1)$. For any vertex $v_i^1 \in s_1$, $v_i^2 \in s_2$, $v_i^3 \in s_3$ and $v_i^4 \in s_4$ the odd graceful labeling functions $f(v_i^1), f(v_i^2), f(v_i^3)$ and $f(v_i^4)$ defined respectively as follows:

$$f(v_i^1) = \begin{cases} i-1 & i \text{ odd} \\ (10n-8)-i & i \text{ even} \end{cases} \qquad f(v_i^2) = \begin{cases} i & i \text{ odd} \\ (12n-11)-i & i \text{ even} \end{cases}$$

$$f(v_i^3) = \begin{cases} 4n+i-5 & i \text{ odd} \\ 6n-i-4 & i \text{ even} \end{cases} \qquad f(v_i^4) = \begin{cases} 2(n-1)+i & i \text{ odd} \\ 6n-i-5 & i \text{ even} \end{cases}$$

The weights of the vertices in the set $s_1$ and $s_3$ are the even numbers 0, 2, 4, …, $12(n-1)-2$ and the weights of the vertices in the set $s_2$ and $s_4$ are the odd numbers 1, 3, 5,…,$12(n-1)-1$. The edge's labeling induced by the absolute value of the difference of the vertex's labeling. The reader can easily find out, that the vertex labels are distinct and all the edge labels are distinct odd labels. Thus, $f$ is odd graceful labeling of $P_n \wedge P_4$. Figure4 shows the method labeling of the graph $P_8 \wedge P_4$ this complete the proof.∎

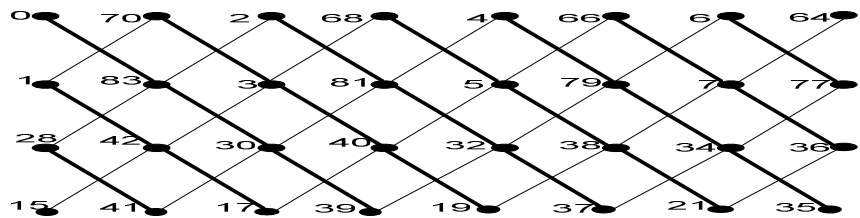

Figure 4 the graph $p_8 \wedge p_4$

### 4.4. Odd gracefulness of $P_n \wedge P_5$

**Theorem 4** $P_n \wedge P_5$ is odd graceful for every integer $n > 2$.

**Proof**

The vertex set of the graph $P_n \wedge P_5$ is partitioned into four disjoint sets $V_1, V_2, V_3, V_4$ and $V_5$ drawn so that the vertices lie on distinct consecutive points of $s_1, s_2, s_3, s_4$ and $s_5$ respectively. The total number of vertices equals $5n$ and total number of edges equals $8(n-1)$. For any vertex $v_i^1 \in s_1, v_i^2 \in s_2, v_i^3 \in s_3, v_i^4 \in s_4$ and $v_i^5 \in s_5$ the odd graceful labeling functions $f(v_i^1), f(v_i^2), f(v_i^3), f(v_i^4)$ and $f(v_i^5)$ defined respectively as follows:

$$f(v_i^1) = \begin{cases} i-1 & i\ odd \\ (14n-12)-i & i\ even \end{cases} \quad f(v_i^2) = \begin{cases} i & i\ odd \\ (16n-15)-i & i\ even \end{cases} \quad f(v_i^3) = \begin{cases} 4n+i-5 & i\ odd \\ 10n-i-8 & i\ even \end{cases}$$

$$f(v_i^4) = \begin{cases} 2n+i-2 & i\ odd \\ 10n-i-9 & i\ even \end{cases} \quad f(v_i^5) = \begin{cases} 8n+i-9 & i\ odd \\ 4(r-3)(n-1)-(2n-4)-i & i\ even \end{cases}$$

The weights of the vertices in the set $s_1, s_3$ and $s_5$ are the even numbers $0, 2, 4, 6, \ldots, 16(n-1)-2$ and the weights of the vertices in the set $s_2$ and $s_4$ are the odd numbers $1, 3, 5, \ldots, 16(n-1)-1$. The edge's labeling induced by the absolute value of the difference of the vertex's labeling. The reader can easily find out, that the vertex labels are distinct and all the edge labels are distinct odd labels. Thus, $f$ is odd graceful labeling of $P_n \wedge P_5$. Figure5 shows the method labeling of the graph $P_8 \wedge P_5$ this complete the proof.∎

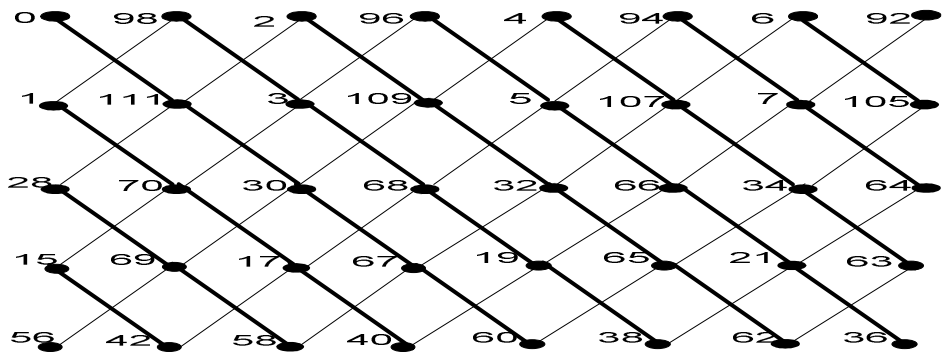

Figure 5 the graph $p_8 \wedge p_5$

**Theorem 5** $P_n \wedge P_6$ is odd graceful for every integer $n > 2$.

**Proof**

The vertex set of the graph $P_n \wedge P_6$ is partitioned into four disjoint sets $V_1, V_2, V_3, V_4, V_5$ and $V_6$ drawn so that the vertices lie on distinct consecutive points of $s_1, s_2, s_3, s_4, s_5$ and $s_6$ respectively. The total number of vertices equals $6n$ and total number of edges equals $10(n-1)$ For any vertex $v_i^1 \in s_1, v_i^2 \in s_2, v_i^3 \in s_3, v_i^4 \in s_4, v_i^5 \in s_5$ and $v_i^6 \in s_6$ the odd graceful labeling function $f(v_i^1)$, $f(v_i^2), f(v_i^3), f(v_i^4), f(v_i^5)$ and $f(v_i^6)$ defined respectively as follows:

$$f(v_i^1) = \begin{cases} i - 1 & i \text{ odd} \\ (18n - 16) - i & i \text{ even} \end{cases} \qquad f(v_i^2) = \begin{cases} i & i \text{ odd} \\ (20n - 19) - i & i \text{ even} \end{cases}$$

$$f(v_i^3) = \begin{cases} 4n + i - 5 & i \text{ odd} \\ 14n - i - 12 & i \text{ even} \end{cases} \qquad f(v_i^4) = \begin{cases} 2(n-1) + i & i \text{ odd} \\ 14n - i - 13 & i \text{ even} \end{cases}$$

$$f(v_i^5) = \begin{cases} 6n + i - 7 & i \text{ odd} \\ 8n - i - 6 & i \text{ even} \end{cases} \qquad f(v_i^6) = \begin{cases} 4n + i - 4 & i \text{ odd} \\ 8n - i - 7 & i \text{ even} \end{cases}$$

The weights of the vertices in the set $s_1, s_3$ and $s_5$ are the even numbers 0, 2, 4, 6,…, 20(n-1)-2 and the weights of the vertices in the set $s_2, s_4$ and $s_6$ are the odd numbers 1, 3, 5,…, 20(n-1)-1. The edge's labeling induced by the absolute value of the difference of the vertex's labeling. The reader can easily find out, that the vertex labels are distinct and all the edge labels are distinct odd labels. Thus, $f$ is odd graceful labeling of $P_n \wedge P_6$. Figure6 shows the method labeling of the graph $P_7 \wedge P_6$ this complete the proof.■

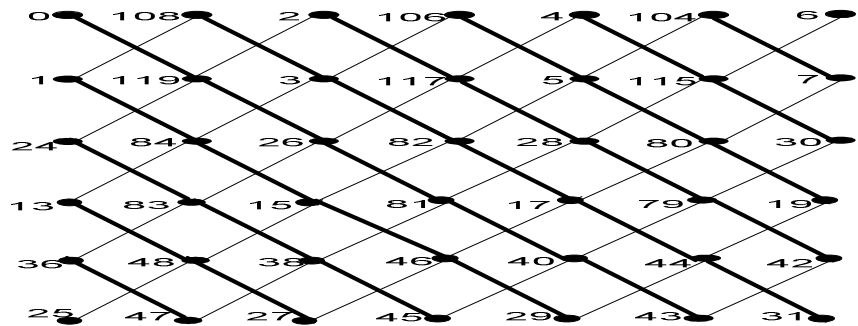

Figure 6 the graph $P_7 \wedge P_6$

### 4.5. The proposed sequential Algorithm

We are introducing an algorithm for the odd graceful labeling of the tensor graph $P_n \wedge P_m$. The algorithm runs in three passes; they and their steps can run in a sequential or a parallel way. In the first pass, the algorithm computes the odd label function for the vertices lie on line $l_1$, see algorithm1 for the initial procedure.

1. Draw the graph $P_n \wedge P_m$ so that the vertices $V_1, V_2, ..., V_m$ lie on distinct consecutive points of $l_1, l_2, ..., l_m$ respectively
2. For each vertex $v_i^1$ on the first level $l_1$ number the vertex as

$$f(v_i^1) = i - 1 \quad (i \text{ odd})$$
$$f(v_i^1) = 4(m-1)(n-1) - 2(n-2) - i \quad (i \text{ even})$$

Algorithm 1: Procedure Initialization

In the second pass, the algorithm labels the vertices on the lines $l_2, l_3, l_4, ... l_m$, it labels the vertices on the line $l_2$ with the labels

$$1, 4m(n-1) - (4n-3), 3, 4m(n-1) - (4n-1), ..., 4(n-1) + (i (i \text{ odd}) - 1), 4(m-1)(n-1) - (2n-4) - i (i \text{ even}), ...,$$

It labels the vertices on the line $l_3$ with the labels

$$4(n-1), 4m(n-1) - 6(n-1), 4n-2, 4m(n-1) - 6n, ..., 4(n-1) + (i (i \text{ odd}) - 1), 4(m-1)(n-1) - (2n-4) - i (i \text{ even}), ...$$

The algorithm traverses to the next lines and it repeats previous processes as shown in algorithm2 until visits all the vertices on the line $l_m$. If $m$ is odd and more than 3; the vertices lie on the line $l_m$ assign values according to step 3 in algorithm 2, otherwise if $m \leq 3$ these vertices assign values according to functions in steps 1,2 in algorithm 2.

1. For each level $l_j$, $j$ is odd and more than three, number its vertices by

   $f(v_i^j) = (j+1)(n-1) + (i-1)$  (i odd)
   
   $f(v_i^j) = 4(m - \lceil j/2 \rceil)(n-1) - ((j-1)n - (j+1)) - i$  (i even)

2. For each level $l_j$, $j$ is even and more than or equal two, number its vertices $v_i^j$ by

   $f(v_i^j) = (j-2)(n-1) + i$  (i odd)
   
   $f(v_i^j) = 4(m - j/2)(n-1) - ((j-2)n - (j-1)) - i$  (i odd)

3. If ($m$ is odd and $m > 3$)

   $f(v_i^m) = \begin{cases} (m+3)n - (m+4) + i & i \text{ odd} \\ 4\lfloor m/2 \rfloor(n-1) - ((m-3)n - (m-1)) - i & i \text{ even} \end{cases}$

### Algorithm2

In the last pass, the label's distribution of the edge between the lines $l_1, l_2, l_3, l_4, \ldots l_m$ is induced by the absolute value of the difference of the vertex's labeling; the following algorithm gives the odd edge labelings.

1. The edge's labels between the line $l_1$ and the line $l_2$, induced by

   $|f^*(v_i^1 v_{i+1}^2)| = 4(m-1)(n-1) - (2i-1)$  (i odd)
   
   $|f^*(v_i^1 v_{i-1}^2)| = 4(m-1)(n-1) - (2i-3)$  (i odd)
   
   $|f^*(v_i^1 v_{i-1}^2)| = 4(m-1)(n-1) - 2(n-2) - (2i-1)$  (i even)

2. The label's distribution of the edge between the lines $l_2, l_3, \ldots l_m$ is defined consecutively by

   $|f^*(v_i^j v_{i+1}^{j+1})| = 2(2m - 2j - 1)(n-1) - (2i-1)$  (i, j odd)
   
   $|f^*(v_i^j v_{i-1}^{j+1})| = 2(2m - 2j - 1)(n-1) - (2i-3)$  (i, j odd)
   
   $|f^*(v_i^j v_{i+1}^{j+1})| = 4(m-j)(n-1) - (2i-1)$  (i, j even)
   
   $|f^*(v_i^j v_{i-1}^{j+1})| = 4(m-j)(n-1) - (2i-3)$  (i, j even)

International journal on applications of graph theory in wireless ad hoc networks and sensor networks
(GRAPH-HOC) Vol.3, No.1, March 2011

3. If ($m$ is odd and $m > 3$

$$|f^*(v_i^{m-1} v_{i+1}^m)| = 4n - 2i - 3 \quad (i \text{ odd})$$

$$|f^*(v_i^{m-1} v_{i-1}^m)| = 4(n-1) - (2i-3) \quad (i \text{ odd})$$

$$|f^*(v_i^{m-1} v_{i+1}^m)| = 2n - 2i - 1 \quad (i \text{ even})$$

$$|f^*(v_i^{m-1} v_{i-1}^m)| = 2n - (2i-1) \quad (i \text{ even})$$

4. Muffle all the edge's labels that contain a vertex not in the vertex set $V(P_n \wedge P_m)$ viz. :

$f^*(v_i^j v_{i+1}^{j+1})$ when $i = n$, and $f^*(v_i^j v_{i-1}^{j+1})$ when $i = 1$, for all $j = 1, 2, ..., m-1$

5. The resulting labeling is odd graceful.

Algorithm 3

**Theorem 6** The tensor graph $P_n \wedge P_m$ is odd graceful for all $n$ and $m$.

**Proof:**

The vertex set of the graph $P_n \wedge P_m$ partitioned into the vertex sets $V_1 = \{v_1^1, v_2^1, ..., v_n^1\}$, $V_2 = \{v_1^2, v_2^2, ..., v_n^2\}$, ... , $V_m = \{v_1^m, v_2^m, ..., v_n^m\}$ lie on distinct consecutive points of $l_1, l_2, ..., l_m$ respectively, with edges only between the vertices in $V_j$ (lie on line $l_j$) and $V_{j+1}$ (lie on line $l_{j+1}$) for all $1 \leq j \leq m-1$, so the graphs $P_n \wedge P_m$ are $m$-partite graphs. The vertices lie on the line $l_1$ are labeled consecutively with even values from the following numbering

$$f(v_i^1) = i - 1, \quad i = 1, 3, 5, ....$$
$$f(v_i^1) = 4(m-1)(n-1) - 2(n-2) - i \quad i = 2, 4, ......$$

The vertices lie on the odd lines $l_j$, $j = 3, 5, ...$ are labeled consecutively with even values from the following numbering

$$f(v_i^j) = (j+1)(n-1) + (i-1), \quad i = 1, 3, 5, ...$$
$$f(v_i^j) = 4(m - \lceil j/2 \rceil)(n-1) - ((j-1)n - (j+1)) - i, \quad i = 2, 4, 6, ...$$

The vertices lie on the even lines $l_j$, $j = 2, 4, 6, ...$ are labeled consecutively with odd values from the following numbering

$$f(v_i^j) = (j-2)(n-1) + i, \quad i = 1, 3, 5, ....$$
$$f(v_i^j) = 4(m - j/2)(n-1) - ((j-2)n - (j-1)) - i, \quad i = 2, 4, 6, .....$$

Maximum and Minimum Values between the resulting vertex labels $f(v_i^j)$ are;

$$f(v_2^2) = \max\left\{\max_{1\leq i\leq n} f(v_i^1),\ \max_{\substack{1\leq i\leq n \\ 1\leq j\leq m}} f(v_i^j),\ \max_{1\leq i\leq n} f(v_i^m)\right\} = 4(n-1)(m-1)-1 = 2q-1$$

$$f(v_1^1) = \min\left\{\min_{1\leq i\leq n} f(v_i^1),\ \min_{\substack{1\leq i\leq n \\ 1\leq j\leq m}} f(v_i^j),\ \min_{1\leq i\leq n} f(v_i^m)\right\} = 0$$

The reader can find out easily that the function $f$ is one – to – one the vertex set of $P_n \wedge P_m$, and the vertex labels are assigned uniquely as shown in algorithm 2. Thus, the labels function $f(v_i^j) \in \{0,1,2,...,2q-1\}$ for all $1\leq i\leq n$, $1\leq j\leq m$. Because the edges only between the vertices in $V_1, V_3, V_5,...$, (are labeled with even values) and in $V_2, V_4, V_6,...$, (are labeled with odd values), then the absolute values of the difference of the vertex's labelings are odd value as shown in algorithm 3. It remains to show that the labels of the edges of $P_n \wedge P_m$ are all the integers of the interval $[1, 2q\text{-}1] = [1, 4(n-1)(m-1)-1]$.

To prove that all edges' labels are different, without loss of generality, suppose that $n$ is even number, we have to consider the following cases

(i) Step 1 of Algorithm 3;

$$\max\left\{\max_{1\leq i\leq n} f^*(v_i^1 v_{i\pm 1}^2)\right\} = 4(n-1)(m-1) - 1 = 2q - 1 \qquad \text{if } i = 1$$

$$\min\left\{\min_{1\leq i\leq n} f^*(v_i^1 v_{i\pm 1}^2)\right\} = 4(n-1)(m-2) + 1 \qquad \text{if } i = n$$

The edges' labels are odd, distinct and are numbered between the above maximum and minimum values according to the decreasing ($i = n-1$ is odd value)

$4(n-1)(m-1)-1,..., 2(n-1)(2m-3)+3, 2(n-1)(2m-3)+1,..., 2(n-1)(2m-3)-3,..., 4(n-1)(m-2)+1$

(ii) Step 2 of Algorithm 3;

$$\max\left\{\max_{\substack{1\leq i\leq n \\ 1\leq j\leq m-1}} f^*(v_i^j v_{i\pm 1}^{j+1})\right\} = f^*(v_2^2 v_1^2) = 4(n-1)(m-2) - 1$$

$$\min\left\{\min_{\substack{1\leq i\leq n \\ 1\leq j\leq m-1}} f^*(v_i^j v_{i\pm 1}^{j+1})\right\} = f^*(v_n^{m-1} v_{n-1}^m) = 1$$

The edges' labels are odd, distinct and are numbered between the above maximum and minimum values according to the decreasing ($i = n-1$ is odd value)

$$4(n-1)(m-2) - 1, 4(n-1)(m-2) - 3, \ldots,\ 2(n-1)(2m-7) - 1, \ldots,\ 2(n-1)(2m-7) - 5, \ldots,\ 3, 1.$$

(iii) Step 3 of Algorithm 3; this step is running only if $n$ is odd number, so it induced a subset of the edge's labeling, which induced in Step 2.

$$\max\left\{\max_{1\leq i\leq n} f^*(v_i^{m-1} v_{i\pm 1}^m)\right\} = \left|f^*(v_1^{m-1} v_2^m)\right| = 4n - 5 = 2q - 1$$

$$\min\left\{\min_{1\leq i\leq n} f^*(v_i^{m-1} v_{i\pm 1}^m)\right\} = \left|f^*(v_{n-1}^{m-1} v_n^m)\right| = 1$$

The edges' labels are odd, distinct and are numbered between the above maximum and minimum values according to the decreasing sequence $4n-5, 4n-7, \ldots,\ 2n-3, 2n-5, \ldots,\ 2n-1, \ldots,\ 3, 1$. We consider this subset $f^*(v_i^{m-1} v_{i\pm 1}^m),\ 1\leq i\leq n$ of the edge's labeling induced in Step 3 if and only if $n$ is odd number, and we consider the subset $f^*(v_i^{m-1} v_{i\pm 1}^m),\ 1\leq i\leq n$ of the edge's labeling induced in Step 2 if and only if $n$ is even number. It is easy to see that in any of the above cases, the edge labels of the graph $P_n \wedge P_m$ for all $m$ and $n$ are all distinct odd integers of the interval $[1, 2q-1] = [1, 4(n-1)(m-1)-1]$ the above algorithm gives an odd gracefulness of $P_n \wedge P_m$ ∎

The algorithm is traversed exactly once for each vertex in the graph $P_n \wedge P_m$, since the number of vertices in the graph equals $nm$ then at most O($nm$) time is spent in total labeling of the vertices and edges, thus the total running time of the algorithm is O($nm$). The parallel algorithm for the odd graceful labeling of the graph $P_n \wedge P_m$, based on the above proposed sequential algorithm is building easily. Since all the above three algorithms 1, 2, and 3 are independent and there is no reason to sort their executing out, so they are to join up parallel in the same time point.

## 5. CONCLUSION

It is desired to have generalized results or results for a whole odd graceful class if possible. But trying to find a general solution, it frequently yields specialized results only. This work presented the generalized solutions to obtain the odd graceful labeling of the graphs obtained by tensor product $P_n \wedge P_m$ of two path graphs. After we introduced an odd graceful labeling of $P_n \wedge P_2, P_n \wedge P_3$, $P_n \wedge P_4$ and $P_n \wedge P_5$, we described a sequential algorithm to label the vertices and the edges of the graph $P_n \wedge P_m$. The sequential algorithm runs in linear with total running time equals O($nm$). The parallel version of the proposed algorithm, as we showed, existed and it is described shortly.